\documentclass{elsarticle}
\usepackage[utf8]{inputenc}
\usepackage[T1]{fontenc}
\AtBeginDocument{%
  }

\usepackage{latexsym}
\usepackage{tabularx}
\usepackage{multirow}
\usepackage{amsfonts}
\usepackage{listings} 
\usepackage{pifont}
\usepackage{array}
\usepackage{tikz}
\usetikzlibrary{tikzmark,calc}
\usepackage{colortbl}
\usepackage{xcolor}
\usepackage{pst-all} 
\usepackage{pst-poly}
\newpsobject{showgrid}{psgrid}{subgriddiv=1,griddots=10,gridlabels=6pt}
\usepackage{graphicx}
\usepackage{fancybox}
\usepackage{wrapfig}
\usepackage{soul}
\newcommand{\con}{\wedge} 
\newcommand{\dis}{\vee} 
\newcommand{\alw}{\Box} 
\newcommand{\imp}{\Rightarrow} 
\newcommand{\som}{\Diamond} 

\let\ab\allowbreak

\usepackage{todonotes}

\newtheorem{claim}{Claim}
\newtheorem{remark}{Remark}
\newtheorem{definition}{Definition}
\graphicspath{{./}{./fig/}{./fig2/}{./fig3/}}
\begin{document}
\title{RE-oriented Model Development with LLM Support and Deduction-based Verification}

\author{Radoslaw Klimek\\
AGH University of Krakow, Poland\\
rklimek@agh.edu.pl}


\begin{abstract}
The requirements engineering (RE) phase is pivotal in developing high-quality software. 
Integrating advanced modelling techniques with large language models (LLMs) and 
formal verification in a logical style can significantly enhance this process. 
We propose a comprehensive framework that focuses on specific 
Unified Modelling Language (UML) diagrams for preliminary system development. 
This framework offers visualisations at various modelling stages and 
seamlessly integrates large language models and logical reasoning engines. 
The behavioural models generated with the assistance of LLMs are 
automatically translated into formal logical specifications. 
Deductive formal verification ensures that logical requirements and 
interrelations between software artefacts are thoroughly addressed. 
Ultimately, 
the framework facilitates the automatic generation of program skeletons, 
streamlining the transition from design to implementation.
\end{abstract}

\maketitle

\section{Introduction}
\label{sec:introduction}

The role of logic in software engineering is rapidly expanding, 
imparting a new quality to software design processes~\cite{Broy-2013}. 
Specifically, deductive reasoning forms the bedrock of verifying the correctness of behavioural models, 
enabling earlier detection of defects within the software lifecycle. 
Recent years have witnessed spectacular developments in automated reasoning systems within 
formal logic~\cite{Gomes-etal-2008}.

The emergence of \emph{Large Language Models} (LLMs) has led to 
transformative advancements across various domains, including software development. 
LLMs have demonstrated significant potential in automating and 
enhancing multiple phases of the software lifecycle~\cite{Fan-etal-2023}.

\emph{Requirements engineering} (RE) constitutes 
the initial and a critical phase of the software development process, 
focusing on the elicitation, identification
and precise documentation of all services to be delivered within 
the context of operational constraints. 
Initially, requirements are highly abstract artefacts. 
A survey of RE tool vendors indicates that requirements elicitation and 
initial modelling are among the most poorly supported activities~\cite{CarrillodeGea-etal-2012}.
As a natural language is often imprecise, 
generating intermediate models helps clarify requirements before code generation.

This paper introduces a formalised methodology and toolset that integrate 
rapid visual modelling with rigorous logical verification within 
a Formal Integrated Development Environment (F-IDE). 
By combining LLMs with deductive reasoning, 
the approach bridges the gap between intuitive requirements modelling and formal verification, ensuring accuracy and consistency before implementation.
Our framework,
called \emph{Logic-oriented Requirements Engineering},
or LoRE+ for short,
and the plus symbol reflecting support for preliminary system modelling,
provides a structured mechanism for system modelling, 
offering a low-entry threshold while maintaining formal rigour. 
By leveraging the generative potential of LLMs alongside deductive reasoning within a F-IDE, 
it enables automatic derivation of correct-by-construction code skeletons that 
align with specified behavioural models, 
thereby enhancing software reliability and maintainability.
The case study presented in this paper validates the practical effectiveness of 
the proposed IDE-centric framework, 
demonstrating its applicability in requirements engineering and early-stage software design. 
The findings substantiate the soundness of 
the approach and underscore its potential for 
broader adoption in AI-assisted IDEs.

The proposed framework emphasises simplicity, flexibility 
and the iterative nature of the development process. 
It bridges the gap between formal methods and practical software development by 
equipping engineers with tools that ensure models which are both correct and 
aligned with requirements prior to coding. 
This integration of LLM support, 
verified through deductive reasoning within IDEs, 
represents a significant step towards more reliable and efficient software engineering practices.

\begin{figure*}[!htb]
\centering
\includegraphics[width=1.0\columnwidth]{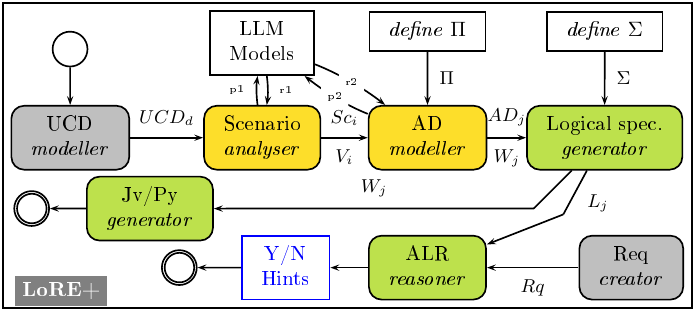}
\caption{The basic system architecture of LoRE+ and its data flows,
including the initial state and final outcomes. 
Grey components represent processes performed by an analyst or area specialist, 
yellow denote automation supervised by an analyst 
and green indicate fully automated processing, 
with iteration possible between phases. 
Interactions (prompts and responses) with LLMs occur at two points within the system. 
The $\Pi$ and $\Sigma$ parameters are defined by an expert once and used widely. 
Inference engines are embedded within the ALR component and are not visualised. 
Code skeletons are generated upon request.
Workflow composition, 
LLM-assisted extraction, and code generation are 
discussed in Sections~\ref{sec:compositional-workflows}--\ref{sec:skeleton-generation}, 
respectively}
\label{fig:architecture}
\end{figure*}

Additionally, 
we consider our approach to be in line with the principle of 
\emph{Correctness-by-Construction} (CbC)~\cite{Runge-etal-2019}, 
which emphasises ensuring correctness through meticulous design, 
verification—including LLMs and deductive reasoning—and the generation of a code compliant with 
requirements during the system's inception phase.

\section{System architecture}

Figure~\ref{fig:architecture} 
illustrates the architecture of the proposed system. 
The system components and data flows are briefly discussed below. 
The \emph{UCD modeller} functions as a conventional tool for creating use case diagrams. 
The generated $UCD_d$ diagram is transmitted to the subsequent component, 
the \emph{Scenario analyser}, 
which models and analyses scenarios $Sc_i$ for each use case. 
During this phase, an LLM is employed to identify, 
within the scenario,
structured yet articulated in the natural language,
elementary, or atomic, system activities. 
With the assistance of the LLM (prompt $p_1$ and response $r_1$), 
the essential vocabulary $V_i$ of activities is thus established,
see also Section~\ref{sec:llm}. 
Subsequently, these activities will also constitute the system's functions, 
see Section~\ref{sec:skeleton-generation}.

The \emph{AD modeller} constructs activity diagrams for individual activities.
Its inputs include both the scenario $Sc_i$ and the vocabulary $V_i$ of identified activities. 
In this phase, LLMs assist (prompt $p_2$ and response $r_2$) 
by proposing activity diagrams $AD_{j=1,\dots}$ as workflows corresponding to 
the scenario $Sc_i$. 
The resulting workflow must be structured and articulated using the $\Pi$ predefined patterns,
see Section~\ref{sec:compositional-workflows}. 
Consequently, 
models proposed by the LLM may require manual modification and alignment to meet this requirement.

Thus, the proposed IDE tool utilises LLMs to automatically identify elementary activities 
within scenarios described in natural language and subsequently recognises their workflows. 
Further details are provided in Section~\ref{sec:llm}.

Subsequent processes are primarily automated. 
The \emph{Logical Specification Generator} 
produces an $L_j$ logical specification equivalent to the activity diagram. 
Its inputs include the designed $AD_j$ activity diagram and
a $W_j$ pattern expression,
which serves as the literal representation of the activity diagram. 
The $\Sigma$ set comprises the necessary logical patterns for 
the logical specification generation process.
This process and approach are discussed in Section~\ref{sec:compositional-workflows}.

The \emph{ALR reasoner} (Automatic Logic Reasoner) 
enables the detection of logical inconsistencies within the constructed behavioural model. 
The reasoning process can be executed either on demand or continuously during regular IDE operations, 
applying deductive inference to behavioural models. 
For instance, 
it can identify contradictory atomic system activities or highlight scenarios 
where the behavioural model fails to satisfy expected properties and constraints defined in logical terms. Consequently, it provides valuable insights to requirements engineering (RE) model designers.

The \emph{Req creator}, also referred to as the \emph{correctness assertions creator}, 
defines the $Rq$ required properties relevant to the behavioural model. 
The outcome, termed ``Y/N (Yes/No) Hints'', 
reflects the result of logical reasoning aimed at providing extensive guidance. 
This is achieved by testing logical formulae that represent both 
the behavioural models and expected assertions, 
using theorem provers to check consistency and constraint satisfaction.
%
Numerous established theorem provers and inference engines are readily available, 
such as E~\cite{Schulz-2002} and Vampire~\cite{Riazanov-Voroknov-2002} for first-order logic, 
and InKreSAT~\cite{Kaminski-Tebbi-2013} for propositional linear temporal logic. 
This work focuses on integrating these tools, 
whose large-scale testing has confirmed their well-established efficiency.

\begin{remark}
The modular architecture of the system presented is both flexible and transparent, 
facilitating the implementation of core planned F-IDE functionalities while ensuring 
the ease of maintenance.
\end{remark}
\begin{remark}
The presented structuring shows a natural progression from general artefacts, through LLM support, to increasingly formal behavioural models, which undergo rigorous logical verification.
\end{remark}

\section{Compositional workflows}
\label{sec:compositional-workflows}

The automatic generation of a logical specification equivalent to 
a system's behavioural model is generally challenging, if not unfeasible. 
Conversely, the the manual production of such a specification,
involving numerous logical formulas,
would be laborious, error-prone and virtually impracticable. 
However, as demonstrated above~\cite{Klimek-2019-LAMP,Klimek-Witek-2024-ASE-RENE}, 
the automatic generation of logical specifications is feasible for 
a certain class of models when activity diagrams are constructed in 
accordance with the \emph{composition principle}, 
characterised by structuralism and hierarchical organisation. 
This approach is outlined below.


A set of \emph{approved (structural) patterns}~$\Pi$ has to be accepted within every design process.
\begin{claim}
A set of patterns proposed in this article covers:
\begin{eqnarray}
\Pi = \{Seq, SeqSeq, Cond, Alt, Para, Loop \}\label{for:approved-patterns}
\end{eqnarray}
where,
the successive patterns illustrate:
the sequence of activities,
a double activity sequence,
the conditional choice of activities,
the conditional choice without the ``else'' branch,
the parallelisation of activities,
processing within a loop.
\end{claim}
\begin{claim}
\emph{Compositionality} in workflows ensures structural coherence through 
a clear partitioning of the system into components, 
including their nesting 
and unambiguous connections, 
resulting in a readable and logically organised structure.
\end{claim}
\begin{definition}
\label{def:workflow-semantics}
The approved set of patterns is defined as follows:
\begin{itemize}
\item $Seq(a1,a2)\equiv a1;a2$,
\item $SeqSeq(a1,a2,a3)\equiv a1;a2;a3$,
\item $Cond(a1,a2,a3,a4)\equiv \mbox{\textbf{if}~} a1 \mbox{~\textbf{then}~} a2 \mbox{~\textbf{else}~} a3; a4$,
\item $Para(a1,a2,a3,a4)\equiv a1; a2||a3; a4$,
\item $Alt(a1,a2,a3)\equiv \mbox{\textbf{if}~} a1 \mbox{~\textbf{then}~} a2; a3$,
\item $Loop(a1,a2,a3,a4)\equiv a1; \mbox{\textbf{while}~} a2 \mbox{~\textbf{do}~} a3; a4$,
\end{itemize}
where $a1$, $a2$, $a3$ and $a4$ represent (atomic) activities, that is pattern formal arguments.
\end{definition}
The introduced patterns,
see Formula~(\ref{for:approved-patterns}),
are slightly redundant in respect of the modelling of arbitrary behaviours.
This redundancy occurs purely for the sake of convenience and carries no underlying intent.
For example,
$Seq$, $Cond$, $Para$ and $Loop$ patterns are fully sufficient
when modelling sequence, concurrency, choice, and
iteration~\cite{Pender-2003}
for the arbitrary activity diagrams of UML.
$SeqSeq$ and $Alt$ patterns are redundant and their task is  modelling:
double sequence and exceptions for UML scenarios, respectively,
and they were introduced in order to facilitate the analyst's work.
The logical properties of predefined patterns from Formula~(\ref{for:approved-patterns})
have been defined here, following the work of~\cite{Klimek-2019-LAMP}, 
in Table~\ref{tab:fixed-properties-patterns-pltl}. 

\begin{table}[!htb]
\caption{A set of fixed $\Sigma$ logical properties for Formula~(\ref{for:approved-patterns})}
\begin{tabularx}{\columnwidth}{|cXc|}
\hline
$\Sigma=\{$ &
$\mathbf{Seq}(a1,a2)=
\langle a1, a2,
\som a1,
\alw(a1 \imp\som a2),\ab 
\alw\neg(a1 \con a2)
\rangle,$ 
& \\
&
$\mathbf{SeqSeq}(a1,a2,a3)=
\langle a1,a3,
\som a1,\ab
\alw(a1\imp \som a2),\ab
\alw(a2\imp \som a3),\ab
\alw\neg(a1 \con a2),\ab
\alw\neg(a2 \con a3)
\rangle,$
&\\
&
$\mathbf{Cond}(a1,a2,a3,a4)=
\langle
a1,
a4,
\som a1,\ab
\alw(a1 \imp (\som a2 \con\neg\som a3) \dis (\neg\som a2 \con \som a3)),\ab
\alw(a1^+ \imp \som a2),\ab
\alw(a1^- \imp \som a3),\ab
\alw(a2 \dis a3 \imp \som a4),\ab
\alw\neg(a1 \con (a2 \dis a3)),\ab
\alw\neg((a2 \dis a3) \con a4)
\rangle,$
&\\
&
$\mathbf{Para}(a1,a2,a3,a4)=
\langle
a1,
a4,
\som a1,\ab
\alw(a1 \imp \som a2 \con \som a3),\ab
\alw(a2 \imp \som a4),
\alw(a3 \imp \som a4),\ab
\alw \neg(a1 \con (a2 \dis a3)),\ab
\alw \neg((a2 \dis a3) \con a4)
\rangle,$
&\\
&
$\mathbf{Alt} (a1,a2,a3)=
\langle
a1,
a3,
\som a1,
\alw(a1^+ \imp \som a2),\ab
\alw(a1^- \imp \neg\som a2 \con \som a3),\ab
\alw(a2 \imp \som a3),\ab
\alw\neg(a1 \con a2),\ab
\alw\neg(a1 \con a3),\ab
\alw\neg(a2 \con a3)
\rangle $,
&\\
%
&
$\mathbf{Loop} (a1,a2,a3,a4)=
\langle
a1,
a4,
\som a1,
\alw(a1 \imp \som a2),\ab
\alw(a2 \imp (\som a3 \con \som a4) \dis (\neg\som a3 \con \som a4)),\ab
\alw(a2 \con a2^+ \imp \som a3),
\alw(a2 \con a2^- \imp \neg\som a3 \con \som a4),\ab
\alw(a3 \imp \som a2),\ab
\alw(a4 \imp \neg\som a2 \con \neg\som a3),\ab
\alw\neg(a1 \con (a2 \dis a3 \dis a4)),\ab
\alw\neg(a2 \con (a1 \dis a3 \dis a4)),\ab
\alw\neg(a3 \con (a1 \dis a2 \dis a4)),\ab
\alw\neg(a4 \con (a1 \dis a2 \dis a3))
\rangle$
& \\
\qquad \} & & \\
\hline
\end{tabularx}
\label{tab:fixed-properties-patterns-pltl}
\end{table}

Generating logical specifications takes place in two steps.
The $\Pi scan$ algorithm for generating a pattern expression
could be written in symbols as:
\begin{eqnarray}
\Pi scan(\Pi,AD_{j}) \longrightarrow W_j\label{for:algorithm-pi-scan}
\end{eqnarray}
where $\Pi$ is a set of approved patterns,
$AD_j$ is a compositionally formed activity diagram and 
$W_{j}$ is an expression literally representing a processed activity diagram,
see, for example, Formula~(\ref{for:expression-ad2}).
The algorithm $\Pi C$~\cite{Klimek-2019-LAMP} for generating logical specification 
could be written in symbols as:
\begin{eqnarray}
\Pi C(\Sigma,W_{j}) \longrightarrow L_j\label{for:algorithm-pi-c}
\end{eqnarray}
where $\Sigma$ is a set of fixed logical properties
and $W_{j}$ is an expression for a processed activity diagram.
$L_j$ is an output set of logical formulas expressed in 
\emph{Propositional Linear-Time Temporal Logic}
(PLTL)~\cite{Emerson-1990},
see, for example, Formula~(\ref{for:generated-specification}),
which is equivalent to the processed activity diagram.

\begin{remark}
The presented approach, 
owing to its rigorous compositional framework, 
enables the automatic generation of a logical specification equivalent to 
a developed, arbitrarily nested workflow, 
thus serving as a behavioural model of the designed system.
\end{remark}

Given that PLTL logic is more expressive than \emph{First-Order Logic} (FOL)~\cite{Kleene-1952}, 
applying our approach to FOL should not present significant challenges. 
(It is possible to consider FOL as 
\emph{minimal temporal logic}~\cite{vanBenthem-1995}.)
This is particularly advantageous, as theorem provers for FOL are more readily available than those for PLTL. Therefore, although the original patterns were defined in PLTL, 
we utilised FOL for testing theorem provers by simplifying the logical patterns through 
the substitution of temporal operators, $\alw$ and $\som$, 
with universal and existential quantifiers. 
In future work, we plan to define logical patterns specifically for FOL.


\section{LLM-powered extraction}
\label{sec:llm}

The process of extracting structured representations from natural language scenarios is 
divided into two distinct identification tasks, 
both utilising the one-shot prompting technique~\cite{Berryman-Ziegler-2024}.
These tasks target the different aspects of scenario interpretation, 
leveraging LLMs to automate and optimise the initial design stages.

The first identification task focuses on extracting atomic activities and 
their associated formal parameters. 
The LLM is prompted, see Listing~\ref{lst:llm-prompt1},
to infer the vocabulary of fundamental activities.
The second identification task involves generating structured activity diagrams 
based on predefined workflow patterns. 
The LLM is prompted, see Listing~\ref{lst:llm-prompt2},
to derive structured representations that conform to compositional principles, 
ensuring seamless integration into the system model.

In our experiments, we utilised ChatGPT-4.o. 
While one could debate on the choice and analyse relevant benchmarks, 
at this stage, we opted for one of the most widely used tools, 
which also maintains a high standard of performance.
Each prompt scaffolding for an LLM consists of 
a \emph{static part} (invariable), 
as presented in Listings~\ref{lst:llm-prompt1} and~\ref{lst:llm-prompt2}, 
and a \emph{dynamic part} (variable), 
which comprises the analysed scenarios.
Each scenario is initially processed through the first query and subsequently, 
along with the identified tags, is passed to the second query.

{\small
\begin{lstlisting}[label={lst:llm-prompt1},caption={Prompt 1 (static part of $p1$ in Fig.~\ref{fig:architecture}): Extraction of activities and parameters}]
For the given use case scenario in natural language, 
please apply the following tags:
- <A>...</A> to denote primary activities, where 
     the entire verb-noun phrase represents 
     a single activity.
- <P>...</P> to denote activity parameters 
    (input data, attributes, or values necessary 
    for performing the activity, if applicable).
Tag short semantically related phrases rather than 
entire sentences. Each primary activity should 
include both the verb and its object as a single 
unit within <A>...</A>. Tags should not 
encompass actors. Example:
The user <A>selects a product</A> and provides 
the <P>product name</P> and <P>quantity</P>.
The system <A>initiates the payment process</A> 
for the <P>product name</P> and <P>quantity</P>.
\end{lstlisting}
}

{\small
\begin{lstlisting}[label={lst:llm-prompt2},caption={Prompt 2 (static part of $p2$ in Fig.~\ref{fig:architecture}): Extraction of fundamental structures}]
For the given use case scenario written in natural 
language, identify and mark control flow structures 
using the following tags:
- <SEQ>...</SEQ> for sequences of activities (steps 
      performed in a specific order), it may contain 
      one or two activities.
- <COND>...</COND> for conditional activities (steps 
      executed based on a condition).
- <ALT>...</ALT> for conditional activities without 
      an 'else' branch.
- <PARA>...</PARA> for parallel activities 
     (operations performed simultaneously).
- <LOOP>...</LOOP> for loops (repeated activities).
Structural tags should enclose activities previously 
marked with <A>...</A>. Nested structures are allowed 
and welcome-apply them logically. Do not tag 
actors-only activities! Example:
The user <SEQ> <A>enters data</A>. The system 
<A>verifies the data</A> and <A>stores it in 
the database</A> </SEQ>. <COND> If <A>the data is 
incorrect</A>, the system <A>sends a notification</A> 
and <A>rejects the order</A> </COND>.
<ALT> If <A>the data is valid</A>, the system 
<A>completes the order</A> </ALT>.
If <A>there are more records</A>, <LOOP> the system 
<A>processes the next record</A> </LOOP>. Meanwhile, 
<PARA> <A>the change history is updated</A> and 
<A>events are logged</A> </PARA>.
\end{lstlisting}
}

\begin{claim}
This formulation of both prompts is the result of 
extensive tuning experiments.
The proposed one-shot LLM-based identification method demonstrates 
remarkable reliability in extracting atomic activities and
requires minimal human intervention in extracting pattern structures.
The approach remains a fundamental tool for 
enabling compositional modelling of activity diagrams.
\end{claim}


\section{Skeleton generation}
\label{sec:skeleton-generation}

Each $W_j$ pattern expression also constitutes a reliable basis for generating source code skeletons. 
A \emph{skeleton} is a high-level program structure containing a dummy code, 
resembling a pseudocode but typically parseable and compilable. 
Skeletons facilitate processing simulation with a designed control flow, 
often including empty function declarations or placeholder functions returning expected results.
To achieve this, 
we define a context-free grammar~\cite{Hopcroft-etal-2006}, 
which serves as input for ANTLR~\cite{ANTLR-tool} 
to generate a syntax checker (parser)~\cite{Hopcroft-etal-2006}. 
By using the \emph{listener} or \emph{visitor} mechanisms~\cite{ANTLR-tool}, 
the system then constructs a generator, 
forming a complete parser-generator pipeline that produces Java skeletons 
for any $W_j$ input expression,
see ``Jv/Py code generator'' in Figure~\ref{fig:architecture}.
We define the following grammar.
\begin{definition}
Grammar production rules intended for the code generation in Java
(the numbers labelling each rule are provided in the square brackets,
treated here as a form of comments,
see also Claim~\ref{cla:grammar-expression}):\\
$\langle$Pattern-rules$\rangle$ $\rightarrow$ [1] $\langle$2-arg-pat$\rangle$ | [2] $\langle$3-arg-pat$\rangle$ | [3] $\langle$4-arg-pat$\rangle$\\
$\langle$2-arg-pat$\rangle$ $\rightarrow$ [4] Seq $\langle$2-args$\rangle$\\
$\langle$3-arg-pat$\rangle$ $\rightarrow$ [5] SeqSeq $\langle$3-args$\rangle$ | [6] Alt $\langle$3-args$\rangle$\\
$\langle$4-arg-pat$\rangle$ $\rightarrow$ [7] Cond $\langle$4-args$\rangle$ | [8] Para $\langle$4-args$\rangle$ | [9] Loop $\langle$4-args$\rangle$\\
$\langle$2-args$\rangle$ $\rightarrow$ [10] ( $\langle$arg$\rangle$ , $\langle$arg$\rangle$ )\\
$\langle$3-args$\rangle$ $\rightarrow$ [11] ( $\langle$arg$\rangle$ , $\langle$arg$\rangle$ , $\langle$arg$\rangle$ )\\
$\langle$4-args$\rangle$ $\rightarrow$ [12] ( $\langle$arg$\rangle$ , $\langle$arg$\rangle$ , $\langle$arg$\rangle$ , $\langle$arg$\rangle$ )\\
$\langle$arg$\rangle$ $\rightarrow$ [13] $\langle$Pattern-rules$\rangle$ | [14] ident($\langle$Java-args$\rangle$)\\
$\langle$Java-args$\rangle$ $\rightarrow$ [15] $\varepsilon$\\
where all non-terminal symbols are enclosed in angle brackets and
$\langle$Pattern-rules$\rangle$ is the grammar start symbol.
\end{definition}
\begin{remark}
The grammar terminal symbols $\{ Seq, SeqSeq, Cond$, $Alt, Para, Loop \}$ are equivalent to 
the approved patterns, 
see~Formula~(\ref{for:approved-patterns}).
\end{remark}

The empty symbol in the last grammar production indicates that in the current initial version 
we do not consider supplying actual arguments to the function, 
see Listing~\ref{lst:Java-AD2}. 
However, we have already conducted experiments where arbitrary arguments, 
following the Java syntax, are supplied. 
This requires a simple extension of the aforementioned production, 
which has been omitted here to avoid obfuscating the overall picture. 
(The same applies to the Python syntax.)
\begin{claim}
The compositional approach,
see Section~\ref{sec:compositional-workflows}, 
entails the pattern expression generation,
see Formula~(\ref{for:algorithm-pi-scan}), 
that literally represents the intended workflow. 
The presented grammar generates a language of pattern expressions,
see Formula~(\ref{for:expression-ad2}).
Consequently, upon defining the formal grammar for expressions, 
it becomes feasible to automatically generate a code in the form of a skeleton,
see~Listing~\ref{lst:Java-AD2}.
\end{claim}

\section{Detailed Case Study}
\label{sec:case-study}

\begin{figure}[!htb]
\centering
\includegraphics[width=.6\columnwidth]{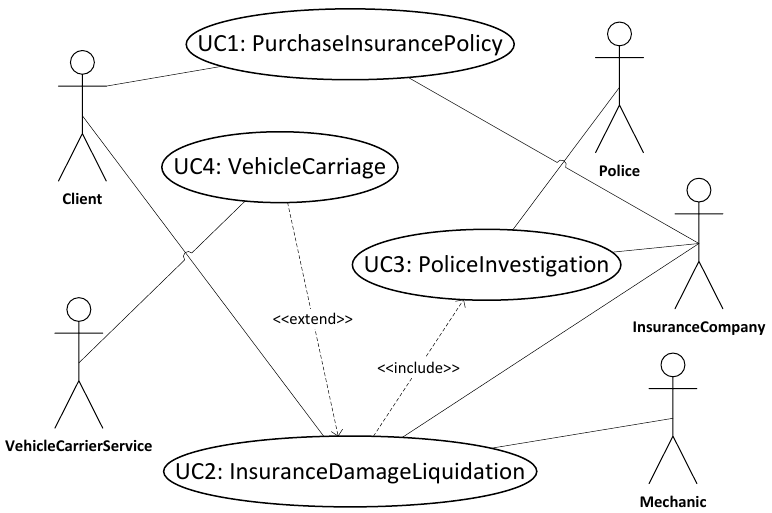}\\
\includegraphics[width=.55\columnwidth]{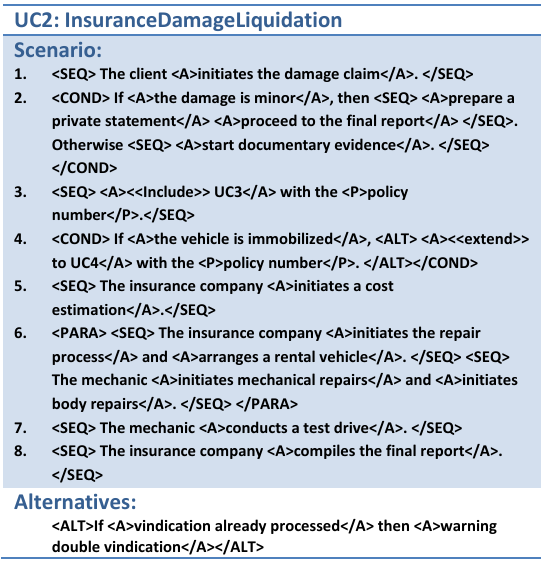}\\
\includegraphics[width=.55\columnwidth]{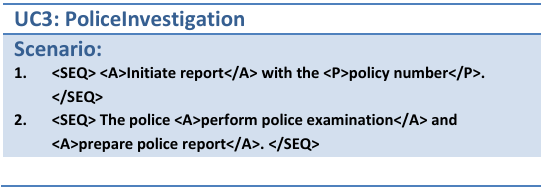}\\
\includegraphics[width=.55\columnwidth]{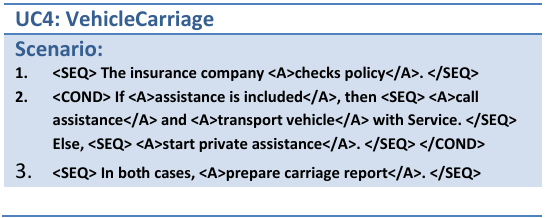}
%
\caption{Use case diagram (top) and 
tagged scenarios for $UC2$, $UC3$, and $UC4$ 
(after prompting, as per Listings~\ref{lst:llm-prompt1}--\ref{lst:llm-prompt2})}
\label{fig:use-case-scenarios}
\end{figure}

\begin{figure}[!htb]
\centering
\includegraphics[width=.6\columnwidth]{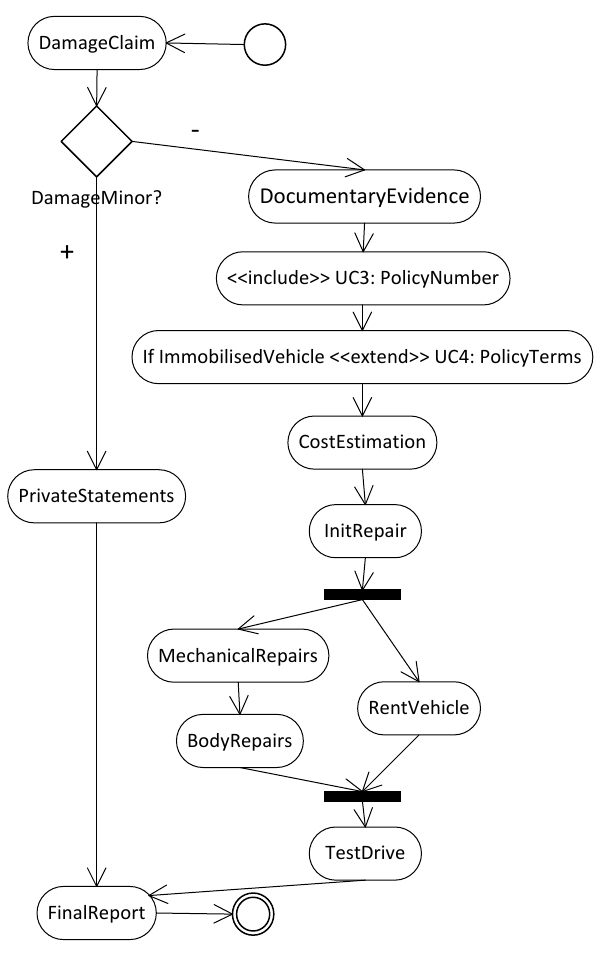}\\
\includegraphics[width=.23\columnwidth]{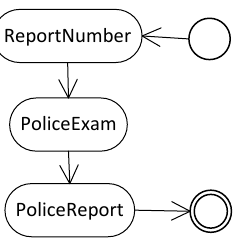}
\includegraphics[width=.35\columnwidth]{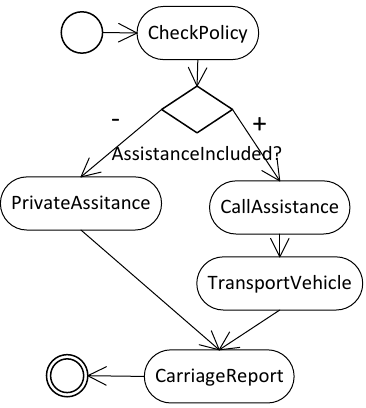}
\caption{Structured activity diagrams $AD_{2}$ (top), 
with $AD_{3}$ and $AD_{4}$ (bottom)
as call behaviour actions, 
with initial and final states}
\label{fig:activity-diagrams}
\end{figure}

Let us assume that we have a use case diagram, $UCD$ ``CarInsurance'',
which consists of several use cases:
$UC1$ ``PurchaseInsurancePolicy'',
$UC2$ ``InsuranceDamageLiquidation'',
$UC3$ ``PoliceInvestigation'' and
$UC4$ ``VehicleCarriage''.
$UC2$ includes $UC3$ and $UC4$ extends $UC2$.
Figures~\ref{fig:use-case-scenarios}
and~\ref{fig:activity-diagrams}
present scenarios and activity diagrams.
The scenarios are appropriately tagged by the LLM in two steps. 
The first step, concerning activities, 
does not raise any significant concerns, 
while the second one provides a solid foundation for analysts to identify and obtain 
the required structural form of the diagram. 
The transition to activity diagrams is guided by pattern-based mapping and supported by human validation in order to minimise potential inaccuracies arising from LLM hallucinations.

\begin{remark}
We observe a natural progression, facilitated by LLMs, 
from narrative, structured scenarios to activity diagrams. 
Within these scenarios, we have identified atomic activities,
complete with parameters, 
and subsequently discerned potential structural patterns.
The application of LLMs enables the significant and beneficial automation of 
this transition, even if subsequent corrections are necessary.
\end{remark}

{\small
(The substitution of numbers as propositions is made,
as a technical matter only,
in places of word propositions,
because of the paper limited size:
$1$ -- DamageClaim,
$2$ -- DamageMinor,
$3$ -- PrivateStatements,
$4$ -- DocumentaryEvidence,
$5$ -- ImmobilisedVehicle,
$6$ -- CostEstimation,
$7$ -- InitRepair,
$8$ -- RentVehicle,
$9$ -- MechanicalRepairs,
$10$ -- BodyRepairs,
$11$ -- TestDrive,
$12$ -- FinalReport,
and
$13$ -- VindicationAlreadyProcessed,
$14$ -- WarningDoubleVindication,
$15$ -- null,
and
$16$ -- ReportNumber,
$17$ -- PoliceExam,
$18$ -- PoliceReport,
and
$19$ -- CheckPolicy,
$20$ -- AssistanceIncluded,
$21$ -- CallAssitance,
$22$ -- TransportVehicle,
$23$ -- PrivateAssistance,
$24$ -- CarriageReport.)
}

The pattern expression for $AD_2$,
including the called activity diagrams $AD_3$ and $AD_4$
(the prime symbol signifies copied variables treated as new atomic variables) is:
{\normalsize
\begin{eqnarray}
W(AD_2)=
Seq(1,Cond(2,3,
SeqSeq(Seq(4,SeqSeq(16',17',\nonumber\\ 18')),
Alt(5,Seq(19,Cond(20',Seq(21',
22'),23',24')),6),\nonumber\\
Para(7,8,Seq(9,10),11)),12))\label{for:expression-ad2}
\end{eqnarray}
}
\begin{claim}
\label{cla:grammar-expression}
The expression in Formula~(\ref{for:expression-ad2}) 
can be derived from the grammar in Definition~\ref{def:workflow-semantics} 
by the sequential application of the following production rules: 
1, 4, 10, 14, 15, 13, 3, 7, 12, 14, 15, 14, 15, 13, 2, 5, 11, 13, 1, 10, 14, 15, 
13, 2, 5, 11, 14, 15, 14, 15, 14, 15, 13, 2, 6, 11, 14, 15, 13, 1, 4, 10, 14, 15,
13, 3, 7, 12, 14, 15, 13, 1, 4, 10, 14, 15, 14, 15, 14, 15, 14, 15, 14, 15, 
13, 3, 8, 12, 14, 15, 14, 15, 13, 1, 4, 10, 14, 15, 14, 15, 14, 15, 14, 15.
\end{claim}
The logical specification generated in accordance with 
Formula~(\ref{for:algorithm-pi-c}), 
given the inputs from Table~\ref{tab:fixed-properties-patterns-pltl} (as $\Sigma$)
and expression Formula~(\ref{for:expression-ad2}), is:
{\small
\begin{eqnarray}
L(AD_2)=\{
%
\som 21',
\alw(21' \imp\som 22'), 
\alw\neg(21' \con 22'),
\som 20',\nonumber\\
\alw(20' \imp (\som (21' \dis 22') \con\neg\som 23') \dis (\neg\som (21' \dis 22') \con \som 23')),\nonumber\\
\alw(20'^+ \imp \som (21' \dis 22')),
\alw(20'^- \imp \som 23'),\nonumber\\
\alw((21' \dis 22') \dis 23' \imp \som 24'),
\alw\neg(20' \con ((21' \dis 22') \dis 23')),\nonumber\\
\alw\neg(((21' \dis 22') \dis 23') \con 24'),
\som 16',
\alw(16'\imp \som 17'),\nonumber\\
\alw(17'\imp \som 18'),
\alw\neg(16' \con 17'),
\alw\neg(17' \con 18'),
\som 19,\nonumber\\
\alw(19 \imp\som (21 \dis 24)), 
\alw\neg(19 \con (21 \dis 24)),
\som 9,
\alw(9 \imp\som 10),\nonumber\\
\alw\neg(9 \con 10),
\som 4,
\alw(4 \imp\som (16' \dis 18')), 
\alw\neg(4 \con (16' \dis 18')),\nonumber\\
\som 5,
\alw(5^+ \imp \som (19 \dis 24')),
\alw(5^- \imp \neg\som (19 \dis 24') \con \som 6),\nonumber\\
\alw((19 \dis 24') \imp \som 6),
\alw\neg(5 \con (19 \dis 24')),
\alw\neg(5 \con 6),\nonumber\\
\alw\neg(2 \con 6),
\som 7,
\alw(7 \imp \som 8 \con \som (9 \dis 10)),
\alw(8 \imp \som 11),\nonumber\\
\alw(3 \imp \som 11),
\alw\neg(7 \con (8 \dis (9 \dis 10))),\nonumber\\
\alw\neg((8 \dis (9 \dis 10)) \con 11),
\som (4 \dis 18'),\nonumber\\
\alw((4 \dis 18')\imp \som (5 \dis 6)),\nonumber\\
\alw((5 \dis 6)\imp \som (7 \dis 11)),
\alw\neg((4 \dis 18') \con (5 \dis 6)),\nonumber\\
\alw\neg((5 \dis 6) \con (7 \dis 11)),
\som 2,\nonumber\\
\alw(2 \imp (\som 3 \con\neg\som (4 \dis 11)) \dis (\neg\som 3 \con \som (4 \dis 11))),\nonumber\\
\alw(2^+ \imp \som 3),
\alw(2^- \imp \som (4 \dis 11)),\nonumber\\
\alw(3 \dis (4 \dis 11) \imp \som 12),
\alw\neg(2 \con (3 \dis (4 \dis 11))),\nonumber\\
\alw\neg((2 \dis (4 \dis 11)) \con 12),
\som 1,
\alw(1 \imp\som (2 \dis 12)), \nonumber\\
\alw\neg(1 \con (2 \dis 12))
\}  \label{for:generated-specification}
\end{eqnarray}
}
\begin{remark}
The manual generation of the logical specification, depicted as Formula~(\ref{for:generated-specification}), 
is challenging and prone to errors. 
The application of the established algorithm to automate this process 
offers a solution to the problem.
\end{remark}

We can test both the satisfiability of the generated specification and 
the adherence to certain 
properties~\cite{Klimek-Szwed-2013-FedCSIS,Klimek-Faber-Kisiel-Dorohinicki-2013-FedCSIS}.
Consider the following minimal set of requirements:
$r_1 \equiv \alw(4 \imp\som 11)$ (\emph{liveness property}~\cite{Alpern-Schneider-1985})
and
$r_2 \equiv \alw\neg (3 \con 21)$ (\emph{safety property}~\cite{Alpern-Schneider-1985}).
\begin{claim}
Every set of logical formulas should describe
both safety and liveness properties of a system, see~\cite{Alpern-Schneider-1985,Manna-Pnueli-1992}.
\end{claim}
Then,
all the properties can be verified separately or together for 
the considered behavioural model:
\begin{equation}
\vdash Cn(L(AD_2)) \imp r_1 \con r_2\label{for:formal-verification}
\end{equation}
where 
\begin{equation}
\{r_1,r_2\}\equiv Rq\label{for:formal-verification=Rq}
\end{equation}
see Figure~\ref{fig:architecture}.
Formula~(\ref{for:formal-verification}) 
contains the conjunction ($Cn$) of all formulas in the generated specification $L(AD_2)$.
This formula also represents an example,
one of many possible input data formulations,
for the ALR reasoner, 
see Figure~\ref{fig:architecture}, 
which employs deductive reasoning to assess the quality of the behavioural model.

And last but not least,
there is the ability to generate a code skeleton for $AD_2$,
see Listing~\ref{lst:Java-AD2}.
(Activities are treated as functions.)
{\small
\begin{lstlisting}[label={lst:Java-AD2},language=Java,caption={The Java skeleton generated for $W(AD_2)$}]
public class SystemAD2 {
   public static void main(String[] args){
   DamageVindication();
   if(MinorDamage()) { PrivateStatements(); } 
   else {
     DocumentaryEvidence();
     ReportNumber();
     PoliceAnalysis();
     PoliceReport();
     if(ImmobilisedVehicle()) {
       PolicyCheck();
       if(AssistanceIncluded()) {
         CallAssitance(); TransportVehicle(); } 
       else { PrivateAssistance(); }
       CarriageReport(); }
     CostEstimation(); InitRepair();
     Thread thread1 = new Thread (new Runnable() {
       @Override public void run() { 
       RentVehicle(); } });
     Thread thread2 = new Thread(new Runnable() {
       @Override public void run() { 
       MechanicalRepairs(); BodyRepairs(); } });
    thread1.start(); thread2.start(); 
    thread1.join(); thread2.join();
    TestDrive(); }
  FinalReport();
  }
public void DamageVindication() { } // Add code here
 ... other functions (as activities) are ...
 ... omitted here due to the paper limited size ...
public void FinalReport() { } // Add code here
\end{lstlisting}
}

\section{Discussion}

Let us summarise the entire approach formally:
\begin{claim}
By starting with the use cases (Figure~\ref{fig:use-case-scenarios}.top),
via scenarios (Figure~\ref{fig:use-case-scenarios}.bottom)
and using LLMs (Listings~\ref{lst:llm-prompt1}--\ref{lst:llm-prompt2}),
we construct structured activity diagrams (Figure~\ref{fig:activity-diagrams}),
basing on predefined patterns (Formula~(\ref{for:approved-patterns})),
and we subsequently generate automatically (Formulas~(\ref{for:algorithm-pi-scan}) and~(\ref{for:algorithm-pi-c}))
the equivalent logical specifications (Formula~(\ref{for:generated-specification}))
for the obtained diagrams.
The entire process involves the conducting of deduction-based 
formal verifications (Formula~(\ref{for:formal-verification})) 
of the designed behavioural models.
Additionally,
based on the received pattern expression (Formula~(\ref{for:expression-ad2})),
we can generate automatically the Java/Python pseudocodes (Listing~\ref{lst:Java-AD2}).
\end{claim}
\begin{claim}
The approach proposed herein is efficient and sound. 
The syntax and semantics of workflows are defined 
(Definition~\ref{def:workflow-semantics}, Table~\ref{tab:fixed-properties-patterns-pltl}). 
The transformation of structural workflows into logical specifications has 
linear time complexity and is sound~\cite{Klimek-2019-LAMP}. 
Logical reasoning (e.g., Formula~(\ref{for:formal-verification})) 
is time-efficient, 
see competition results for provers~\cite{Sutcliffe-2021,Heule-etal-competitions}. 
The pattern expression (e.g., Formula(\ref{for:expression-ad2})) 
is essentially a regular expression~\cite{Hopcroft-etal-2006} 
and can be generated by a context-free grammar~\cite{Hopcroft-etal-2006},
see Claim~\ref{cla:grammar-expression}. 
Parsing and code generation for it have linear complexity and are 
sound~\cite{Hopcroft-etal-2006,Aho-etal-2006}.
\end{claim}

Our approach has many advantages:
\begin{itemize}
\item
Increased automation --
the framework introduces multi-level automation, from scenario analysis and LLM-driven activity diagram generation to automated logical specification extraction and deductive reasoning. By providing real-time AI-driven feedback within the IDE, the system dynamically assists engineers, allowing them to focus on complex design challenges.
\item 
Consistency and formal verification -- 
automated verification mechanisms ensure consistent model generation while minimising human errors. 
The system cross-validates LLM-generated outputs against formal logical rules, 
reducing ambiguity and reinforcing the alignment of models with initial requirements and specifications.
\item
Advancements in generative AI and interactive refinement -- 
the integration of LLMs accelerates workflow generation and improves structural identification in scenarios. 
The framework enables real-time refinements, 
ensuring that engineers receive context-aware recommendations while iterating on models. 
While some manual refinements may be needed, 
these models provide a solid foundation for subsequent deductive verification.
\item
LLM-based extractions --
the experiments confirm the effectiveness of LLM-based prompting, 
though further tuning could enhance consistency and accuracy. 
A promising direction for improvement is \emph{chain-of-thought prompting}~\cite{Wei-etal-2022}.
\item
Robust logical foundations and deductive bridging --
the framework bridges generative AI capabilities with deductive reasoning techniques, 
ensuring that LLM-generated artefacts remain rigorous and verifiable within 
a well-established logical framework~\cite{Broy-2013,Clarke-Wing-etal-1996,DeMol-Primiero-2015}.
\item
Alignment with the CbC paradigm~\cite{Runge-etal-2019} -- 
the fusion of LLM-powered modelling and formal reasoning supports 
Correct\-ness-by-Construction (CbC) principles~\cite{Runge-etal-2019}, 
ensuring that generated models are provably correct before implementation. 
This structured approach minimises costly late-stage errors and improves software reliability.
\item
Enhanced accessibility and developer interaction -- 
the lightweight design of the framework ensures seamless adoption within IDEs, 
allowing developers of varying expertise levels to interact dynamically with AI-assisted prompts and refinements. 
By providing real-time syntax-aware code skeleton generation and error detection mechanisms, 
the framework significantly lowers the entry threshold for formal modelling and logical verification.
\item
Transparent goals and automated development pipelines -- 
the framework system facilitates end-to-end automation, 
from early-stage requirements engineering to automatically verified models and structured code skeletons. 
The system’s ability to dynamically guide users through model adjustments ensures that 
software artefacts evolve in a controlled and structured manner, leading to higher-quality final implementations.
\end{itemize}

\section{Related works}
\label{sec:related-works}


Hurlbut's comprehensive survey~\cite{Hurlbut-1997} 
highlights challenges in formalising UML use cases, 
particularly the transition from unstructured narratives to semi-structured scripts, 
underscoring the informal nature of scenario documentation and validation difficulties. 
Barrett et al.~\cite{Barrett-etal-2009} 
propose a formal syntax and semantics for use cases; 
nevertheless, their approach is more merging-oriented, 
whereas ours focuses on identification. 
Westergaard~\cite{Westergaard-2011} 
translates declarative workflow languages into linear temporal logic, 
which may not align with activity diagrams central to our study. 
Ara{\'u}jo and Moreira~\cite{Araujo-Moreira-2005} 
offer transformation rules from activity diagrams to temporal logic expressions, 
but these are limited to specific diagram structures, 
whereas our method addresses arbitrary, 
nested structures with a more formalised transformation process. 
Yue et al.~\cite{Yue-etal-2015} 
translate natural language into behavioural models, 
including activity diagrams, using a linguistically driven approach. 
In contrast, 
we build robust logical specifications from the outset, 
classify activity diagrams 
and employ a composition-driven method based on behavioural patterns, 
facilitating complexity management and formal verification via theorem provers and solvers. 
Spichkova et al.~\cite{Spichkova-etal-2012} 
advocate for pervasive formal development, 
primarily for embedded and safety-critical systems, using model checking. 
Our approach, however, supports diverse system designs, 
constructs behavioural models using UML during the RE phase
and relies on classical deductive reasoning with integrated provers.

The existing tools, like Coq and Isabelle, 
assist in formal specification and proof construction but 
require expertise in complex proof languages. 
Modelling tools, such as Enterprise Architect and Rational Software Architect, 
offer limited formal verification capabilities. 
Formal verification environments, like SPIN and NuSMV, employ model checking techniques. 
AutoFocus combines model creation with model checking. 
Our IDE, grounded in UML models, utilises deductive reasoning for verification and 
enables code skeleton generation, 
integrating advanced SAT solvers and theorem provers to enhance system capabilities.

Integrating our deductive approach into the Polarsys framework, 
using tools like Capella or Papyrus, 
could support proofs of logical correctness in 
system models—vital for critical applications. 
This would improve reliability and safety. 
Moreover, methods such as 
Autili et al.'s~\cite{Autili-et-al-2007} 
graphical temporal property specifications could ease 
the definition of complex behaviours and support temporal verification.

\section{Conclusions and future works}
\label{sec:conclusion}

This study presents a framework integrating advanced modelling techniques, LLMs, and deduction-based verification to enhance requirements engineering. By leveraging UML diagrams, LLM-assisted behavioural models are automatically translated into formal specifications, ensuring rigorous verification. The approach also enables automatic program skeleton generation, streamlining the transition from design to implementation.

Further improvements include fine-tuning LLM prompting for better accuracy, with chain-of-thought prompting~\cite{Wei-etal-2022} 
as a promising approach. 
Expanding the framework to new domains via domain-specific ontologies could enhance applicability. 
Additionally, reinforcement learning for prompt optimisation may further improve automation and reliability.


The ongoing development of our LLM-powered IDE focuses on fully integrating automated reasoning to support real-time in a logical style validation, scenario refinement, and code generation. Significant progress has already been made in embedding these capabilities within the development workflow, enhancing both efficiency and usability.
As we continue building the IDE, we are also improving interactive AI support by refining AI-driven feedback loops, leveraging adaptive prompting, and incorporating empirical evaluations. These enhancements aim to provide context-aware recommendations, ensuring a more intelligent and responsive development environment.

\bibliographystyle{plain}
\bibliography{bib-rk,bib-rk-main,bib-rk-tools}

\begin{thebibliography}{10}

\bibitem{Aho-etal-2006}
Alfred~V. Aho, Monica~S. Lam, Ravi Sethi, and Jeffrey~D. Ullman.
\newblock {\em Compilers: Principles, Techniques, and Tools (2nd Edition)}.
\newblock {Addison Wesley}, 2006.

\bibitem{Alpern-Schneider-1985}
Bowen Alpern and Fred~B. Schneider.
\newblock Defining liveness.
\newblock {\em Information Processing Letters}, 21 (4):181--185, 1985.

\bibitem{ANTLR-tool}
{ANTLR Development Team}.
\newblock Website for {ANTLR}, 2023.
\newblock accessed on 17-Apr-2023.

\bibitem{Araujo-Moreira-2005}
Joao Araujo and Ana Moreira.
\newblock Integrating uml activity diagrams with temporal logic expressions.
\newblock In {\em Proceedings of 10th International Workshop on Exploring
  Modeling Methods in Systems Analysis and Design (EMMSAD 2005), June 13--14,
  2005, Porto, Portugal}, pages 477--484, 2005.

\bibitem{Autili-et-al-2007}
M.~Autili, P.~Inverardi, and P.~Pelliccione.
\newblock Graphical scenarios for specifying temporal properties: an automated
  approach.
\newblock {\em Automated Software Engineering.}, 14(3):293–340, sep 2007.

\bibitem{Barrett-etal-2009}
Stephen Barrett, Daniel Sinnig, Patrice Chalin, and Greg Butler.
\newblock Merging of use case models: Semantic foundations.
\newblock In {\em 3rd IEEE International Symposium on Theoretical Aspects of
  Software Engineering (TASE'09)}, pages 182--189, 2009.

\bibitem{Berryman-Ziegler-2024}
J.~Berryman and A.~Ziegler.
\newblock {\em Prompt Engineering for LLMs: The Art and Science of Building
  Large Language Model--Based Applications}.
\newblock O'Reilly Media, 2024.

\bibitem{Broy-2013}
Manfred Broy.
\newblock On the role of logic and algebra in software engineering.
\newblock In Peter Paule, editor, {\em Mathematics, Computer Science and Logic
  -- A Never Ending Story: The Bruno Buchberger Festschrift}, pages 51--68.
  Springer International Publishing, 2013.

\bibitem{CarrillodeGea-etal-2012}
Juan~M. Carrillo~de Gea, Joaqu{\'i}n Nicol{\'a}s, Jos{\'e} L.~Fern{\'a}ndez
  Alem{\'a}n, Ambrosio Toval, Christof Ebert, and Aurora Vizca{\'i}no.
\newblock Requirements engineering tools: Capabilities, survey and assessment.
\newblock {\em Information and Software Technology}, 54(10):1142--1157, 2012.

\bibitem{Clarke-Wing-etal-1996}
E.M. Clarke, J.M. Wing, and et~al.
\newblock Formal methods: State of the art and future directions.
\newblock {\em ACM Computing Surveys}, 28 (4):626--643, 1996.

\bibitem{Emerson-1990}
Ernest~Allen Emerson.
\newblock Temporal and modal logic.
\newblock In Jan van Leeuwen, editor, {\em Handbook of Theoretical Computer
  Science}, volume~B, pages 995--1072. MIT Press, 1990.

\bibitem{Fan-etal-2023}
Angela Fan, Beliz Gokkaya, Mark Harman, Mitya Lyubarskiy, Shubho Sengupta, Shin
  Yoo, and Jie~M. Zhang.
\newblock { Large Language Models for Software Engineering: Survey and Open
  Problems }.
\newblock In {\em 2023 IEEE/ACM International Conference on Software
  Engineering: Future of Software Engineering (ICSE-FoSE)}, pages 31--53, Los
  Alamitos, CA, USA, May 2023. IEEE Computer Society.

\bibitem{Gomes-etal-2008}
Carla~P. Gomes, Henry~A. Kautz, Ashish Sabharwal, and Bart Selman.
\newblock Satisfiability solvers.
\newblock In Frank van Harmelen, Vladimir Lifschitz, and Bruce~W. Porter,
  editors, {\em Handbook of Knowledge Representation}, volume~3 of {\em
  Foundations of Artificial Intelligence}, pages 89--134. Elsevier, 2008.

\bibitem{Heule-etal-competitions}
Marijn Heule, Matti J{\"a}rvisalo, and Martin Suda.
\newblock The international {SAT} competitions, web page.
  \texttt{http://www.satcompetition.org/}, 2022.
\newblock accessed on 16-May-2022.

\bibitem{Hopcroft-etal-2006}
John~E. Hopcroft, Rajeev Motwani, and Jeffrey~D. Ullman.
\newblock {\em Introduction to Automata Theory, Languages, and Computation}.
\newblock Addison-Wesley, 2006.

\bibitem{Hurlbut-1997}
Russell~R. Hurlbut.
\newblock A survey of approaches for describing and formalizing use cases.
\newblock Technical Report XPT-TR-97-03, Expertech, Ltd., 1997.

\bibitem{Kaminski-Tebbi-2013}
Mark Kaminski and Tobias Tebbi.
\newblock {InKreSAT}: Modal reasoning via incremental reduction to {SAT}.
\newblock In Maria~Paola Bonacina, editor, {\em 24th International Conference
  on Automated Deduction (CADE 2013), Lake Placid, New York, 9--14 June 2013},
  volume 7898 of {\em Lecture Notes in Computer Science}, pages 436--442.
  Springer, Jun 2013.

\bibitem{Kleene-1952}
Stephen~Cole Kleene.
\newblock {\em Introduction to Metamathematics}.
\newblock Bibliotheca Mathematica. North-Holland, 1952.

\bibitem{Klimek-2019-LAMP}
Rados{\l}aw Klimek.
\newblock Pattern-based and composition-driven automatic generation of logical
  specifications for workflow-oriented software models.
\newblock {\em Journal of Logical and Algebraic Methods in Programming},
  104:201--226, 2019.

\bibitem{Klimek-Faber-Kisiel-Dorohinicki-2013-FedCSIS}
Rados{\l}aw Klimek, {\L}ukasz Faber, and Marek Kisiel-Dorohinicki.
\newblock Verifying data integration agents with deduction-based models.
\newblock In {\em Proceedings of Federated Conference on Computer Science and
  Information Systems ({FedCSIS} 2013), 8--11 September 2013, Krak\'{o}w,
  Poland}, pages 1049--1055. IEEE Xplore Digital Library, 2013.

\bibitem{Klimek-Szwed-2013-FedCSIS}
Rados{\l}aw Klimek and Piotr Szwed.
\newblock Verification of archimate process specifications based on deductive
  temporal reasoning.
\newblock In {\em Proceedings of Federated Conference on Computer Science and
  Information Systems ({FedCSIS} 2013), 8--11 September 2013, Krak\'{o}w,
  Poland}, pages 1131--1138. IEEE Xplore Digital Library, 2013.

\bibitem{Klimek-Witek-2024-ASE-RENE}
Radoslaw Klimek and Julia Witek.
\newblock Automatic generation of logical specifications for behavioural
  models.
\newblock In {\em Proceedings of the 39th IEEE/ACM International Conference on
  Automated Software Engineering Workshops}, ASEW'24, pages 1--7, New York, NY,
  USA, 2024. Association for Computing Machinery.

\bibitem{Manna-Pnueli-1992}
Zohar Manna and Amir Pnueli.
\newblock {\em The Temporal Logic of Reactive and Concurrent Systems --
  Specification}.
\newblock Springer-Verlag New York, Inc., 1992.

\bibitem{DeMol-Primiero-2015}
Liesbeth~De Mol and Giuseppe Primiero.
\newblock When logic meets engineering: Introduction to logical issues in the
  history and philosophy of computer science.
\newblock {\em History and Philosophy of Logic}, 36(3):195--204, 2015.

\bibitem{Pender-2003}
Tom Pender.
\newblock {\em UML Bible}.
\newblock John Wiley \& Sons, 2003.

\bibitem{Riazanov-Voroknov-2002}
Alexandre Riazanov and Andrei Voronkov.
\newblock The design and implementation of vampire.
\newblock {\em AI Commun.}, 15(2,3):91–110, aug 2002.

\bibitem{Runge-etal-2019}
Tobias Runge, Ina Schaefer, Loek Cleophas, Thomas Th{\"u}m, Derrick Kourie, and
  Bruce~W. Watson.
\newblock Tool support for correctness-by-construction.
\newblock In Reiner H{\"a}hnle and Wil van~der Aalst, editors, {\em Fundamental
  Approaches to Software Engineering}, pages 25--42, Cham, 2019. Springer
  International Publishing.

\bibitem{Schulz-2002}
Stephan Schulz.
\newblock E -- a brainiac theorem prover.
\newblock {\em Journal of AI Communications}, 15(2,3):111–126, aug 2002.

\bibitem{Spichkova-etal-2012}
Maria Spichkova, Florian H{\"{o}}lzl, and David Trachtenherz.
\newblock Verified system development with the autofocus tool chain.
\newblock In C{\'{e}}sar Andr{\'{e}}s and Luis Llana, editors, {\em Proceedings
  2nd Workshop on Formal Methods in the Development of Software, {WS-FMDS}
  2012, Paris, France, August 28, 2012}, volume~86 of {\em {EPTCS}}, pages
  17--24, 2012.

\bibitem{Sutcliffe-2021}
Geoff Sutcliffe.
\newblock The 10th ijcar automated theorem proving system competition --
  casc-j10.
\newblock {\em AI Commun.}, 34(2):163–177, jan 2021.

\bibitem{vanBenthem-1995}
Johan van Benthem.
\newblock {\em Handbook of Logic in Artificial Intelligence and Logic
  Programming}, chapter Temporal Logic, pages 241--350.
\newblock 4. Clarendon Press, 1993--95.

\bibitem{Wei-etal-2022}
Jason Wei, Xuezhi Wang, Dale Schuurmans, Maarten Bosma, Brian Ichter, Fei Xia,
  Ed~H. Chi, Quoc~V. Le, and Denny Zhou.
\newblock Chain-of-thought prompting elicits reasoning in large language
  models.
\newblock In {\em Proceedings of the 36th International Conference on Neural
  Information Processing Systems}, NIPS '22, pages 24824--24837. Curran
  Associates Inc., 2022.

\bibitem{Westergaard-2011}
Michael Westergaard.
\newblock Better algorithms for analyzing and enacting declarative workflow
  languages using {LTL}.
\newblock In Stefanie Rinderle{-}Ma, Farouk Toumani, and Karsten Wolf, editors,
  {\em 9th International Conference on Business Process Management (BPM 2011),
  August 28th -- September 2nd 2011, Clermont-Ferrand, France}, volume 6896 of
  {\em Lecture Notes in Computer Science}, pages 83--98. Springer, 2011.

\bibitem{Yue-etal-2015}
Tao Yue, Lionel~C. Briand, and Yvan Labiche.
\newblock {aToucan}: An automated framework to derive {UML} analysis models
  from use case models.
\newblock {\em ACM Transactions on Software Engineering and Methodology},
  24(3):13:1--13:52, May 2015.

\end{thebibliography}
\end{document}